\documentclass[12pt,notoc]{JHEP3}

\usepackage{amsmath,amssymb,euscript,array,cite}

\input{epsf}
\usepackage{epsfig}

\newcommand {\slsh} [1] {\not{\hbox{\kern-2pt${#1}$}}}

\newcommand {\beq} {\begin{equation}}
\newcommand {\eeq} {\end{equation}}
\newcommand {\ber}{\begin{eqnarray*}}
\newcommand {\eer} {\end{eqnarray*}}
\newcommand {\bea}{\begin{eqnarray}}
\newcommand {\eea} {\end{eqnarray}}
\newcommand{\Nfour} {${\cal N}=4\ $}

\newcommand{\Dslash}{\,{\raise.15ex\hbox{/}\mkern-12mu D}}

\title{Anomalous Dimensions from a Spinning D5-Brane}
\author{Adi Armoni \\
Department of Physics,\\ University of Wales Swansea,\\
Swansea, SA2 8PP, UK.\\
E-mail: {\tt a.armoni@swan.ac.uk}}
\preprint{SWAT/06/500}
\abstract{We consider the anomalous dimension of a certain 
twist two operator in \Nfour super Yang-Mills theory. At strong coupling and large-$N$ it is captured by the classical dynamics of a spinning
D5-brane. The present calculation generalizes the result of Gubser, Klebanov and Polyakov (hep-th/0204051): in order to calculate the anomalous dimension of a bound state of $k$ coincident strings, the spinning closed string is replaced by a spinning D5 brane that wraps an $S^4$ inside the $S^5$ part of the $AdS_5 \times S^5$ metric. }

\begin{document}

\section{Introduction}

Twist two operators of the form ${\rm tr}\, \Phi \nabla _+ ^J \Phi$ (where $\Phi$ is either a quark or a gluon and the subscript $+$ refers to a projection onto the light-cone) play an important role in deep inelastic scattering \cite{Callan:1973pu,Parisi:1973nx,Gross:1974cs}. The calculation of anomalous dimensions of such operators at strong coupling is of great interest.

A while ago Gubser, Klebanov and Polyakov \cite{Gubser:2002tv} considered the following operator in \Nfour SYM
\beq
{\cal O} \equiv {\rm tr}\, \Phi \nabla _+ ^J \Phi \label{GKPoperator}
\eeq
($\Phi \equiv \Phi^a t^a$, and $\nabla _ + \equiv \partial _+ + [A_+ ^ a t^a,\cdot]$). They argued that the anomalous dimension of \eqref{GKPoperator} is captured by the semi-classical dynamics of a fast spinning closed string in $AdS_5$. Their result, valid at large-$N$, $\lambda \gg 1$ and $J \gg \sqrt \lambda $, is
\beq
\gamma = \Delta - J = {\sqrt \lambda \over \pi} \log J \, .
\eeq
In this short note we consider a certain generalization of the above analysis. Let us start with the following operator
\beq
{\cal O}_R \equiv {\rm tr}\, (\Phi ^a T^a _{R}) \nabla _+ ^J (\Phi ^a T^a _{R}) \label{Roperator}
\eeq
with 
\beq
\nabla _+ \equiv \partial _+ + [A_+ ^a T^a _{R},\,\cdot\,] \, . 
\eeq
Thus instead of considering an operator with generators in the fundamental of $SU(N)$ we consider an operator with generators in a representation $R$. We are interested in the anomalous dimension of this operator at strong coupling. Let us assume that the $N$-ality of the representation is $k$. In this case the closed spinning string should be replaced by a bound state of $k$ coincident folded spinning strings. These $k$-strings can be 'glued' together in various ways, spanned by the tensor product of the fundamental representation.

 We expect the following
 expression for the anomalous dimension of \eqref{Roperator} \cite{Korchemsky:1988si}

\beq
\gamma _R= \Delta _R - J = f_R(g ^2 _{\rm YM}, N, k) \log J \, .
\eeq
At strong coupling, large $N$, and when $k/N$ is fixed we expect
\beq
\gamma _R=\Delta _R - J = \sqrt \lambda N \tilde f_R(k/N) \log J \, .
\eeq 
We will argue shortly that when $R$ is the antisymmetric representation
 and in the limit $N\rightarrow \infty$, $k/N$ fixed, the $k$-strings bound state is better described as a spinning D5-brane with k units of electric field on its worldvolume. At strong coupling, large $N$, $k/N$ is fixed (but arbitrary) and $J \gg N \sqrt \lambda$ we find
\beq
\gamma _{\rm antisymmetric}=\Delta _{\rm antisymmetric} - J = {2N\sqrt \lambda \over 3\pi ^2} \sin ^3 \theta _k \log J \, , \label{result1}
\eeq
where
\beq
\pi \left ({k\over N} -1 \right ) = {1\over 2}\sin 2\theta _k - \theta _k . \label{result2}
\eeq

The above results are due to a recent progress in calculations of expectation values of Wilson loop operators in \Nfour SYM. Drukker and Fiol \cite{Drukker:2005kx} argued that a Wilson loop with matter in the symmetric representation is better described as a D3-brane with k units of electric flux. Following their work, it was shown \cite{Hartnoll:2006hr,Yamaguchi:2006tq,Gomis:2006sb,Rodriguez-Gomez:2006zz,Hartnoll:2006ib} that a Wilson loop with matter in the antisymmetric representation is better described by a D5 brane whose worldvolume is a minimal surface in the AdS part of the geometry times an $S^4$ inside the $S^5$. A detailed discussion of {\em why} the D5 brane corresponds to a Wilson loop with matter in the antisymmetric representation is given by ref.\cite{Gomis:2006sb}. The identification is supported by a matrix model calculation \cite{Okuyama:2006jc,Hartnoll:2006is} which is valid at arbitrary coupling and recovers the known results for Wilson loops with antisymmetric matter at both weak and strong couplings. Thus whenever a fundamental string, which represents a Wilson loop in the fundamental representation, is replaced by a D5 brane we obtain the antisymmetric representation instead of the fundamental. This argument justifies the identification of the correspondence between the state ${\cal O}_R | 0 \rangle $ (with $R$ the antisymmetric representation) and the spinning D5 brane. A similar argument in favor of the identification \eqref{Roperator} is given at the end of section 3.

The organization of this paper is as follows: in section 2 we derive our main result by considering a spinning D5 brane in AdS space, in section 3 we 're-derive' the same result by considering Wilson loops with cusps. Finally in section 4 we discuss our result.  

\section{Spinning D5-brane}

Consider a D5 brane in $AdS_5 \times S^5$ background. The D5 wraps an $S^4$ inside the $S^5$. The $S^4$ resides at an angle $\theta$ inside the $S^5$ which will be determined dynamically. The two other 'worldsheet' coordinates are exactly as in the case of the spinning string.

It is thus convenient to parametrize the $AdS_5 \times S^5$ metric as follows
\beq
ds^2 = -\left (1+{r^2 \over R^2} \right ) dt^2 + {dr^2 \over {(1 + {r^2 \over R^2})}} + r^2 (d\chi^2 + \sin ^2 \chi d\phi ^2 + \cos ^2 \chi d\psi ^2) +R^2 d\theta ^2 + R^2 \sin ^2 \theta d \Omega _4 ^2
\eeq

Let us start with the D5 motion inside the AdS part of the metric. As already mentioned, the D5 behaves as a closed spinning string. In that case the motion is described by the Nambu-Goto action (we use the notation of \cite{Armoni:2002xp})

\beq 
 S_F = {1\over 2\pi \alpha' } \int d\tau d\sigma \sqrt {{\rm det}\, ^\star g} = 
 {1\over 2\pi \alpha' } \int d\tau d\sigma {dr \over d\sigma} \sqrt {(1 - {\omega ^2 r^2 \over {1 + {r^2 \over R^2}}})} \, ,
\eeq 
where $\omega = {d\phi \over dt}$ is the angular velocity.

The D5-brane action is (from now on we follow the convention of \cite{Hartnoll:2006ib})
\beq
S_{D5} = T_5 \int d\tau d^5 \sigma e ^{-\Phi} \sqrt {{\rm det}\, (^\star g + 2\pi \alpha ' F)}  - ig_s T_5 \int 2\pi \alpha ' F\wedge ^\star C_4 \,.
\eeq  
The relevant part of the four form is $C_4 = {R^4 \over g_s} \left ( {3(\theta - \pi) \over 2} - \sin ^3 \theta \cos \theta - {3\over 2} \cos \theta \sin \theta \right ) {\rm Vol} \, S^4$. In addition we will have an electric worldvolume field $F_{\tau \sigma}$ such that $\delta S / \delta F_{\tau \sigma} = ik$.

After the above insertions are made, the D5 action takes the form
\beq
S_{D5}= {N \sqrt \lambda \over 3\pi^2} \int d\tau d\sigma \left ( \sin ^4 \theta  \sqrt {{\rm det}\, ^\star g - F^2 } + h(\theta) F \right ) \, , \label{action}
\eeq
 where 
\beq
{\rm det} \, ^\star g \equiv  \left  ({dr \over d\sigma } \right )^2 \left (1- {\omega ^2 r^2 \over 1+ {r^2 \over R^2}} \right ) 
\eeq
 and
\beq
h(\theta)=  {3(\theta - \pi) \over 2} - \sin ^3 \theta \cos \theta - {3\over 2} \cos \theta \sin \theta 
\eeq
\vspace{1cm}
The variation of the action \eqref{action} with respect to $F$ yields the relation \eqref{result2}.
 Variation with respect to $\theta$ yields $F=-\sqrt {{\rm det}\, ^\star g} \cos \theta $ and hence  
\beq
S_{D5}= {N \sqrt \lambda \over 3\pi^2} \int d\tau d\sigma \sqrt {{\rm det}\, ^\star g} \left ( \sin ^5 \theta - h(\theta) \cos \theta \right) \, , \label{afinal}
\eeq

As explained in ref.\cite{Drukker:2005kx} we need to add to \eqref{afinal} boundary terms. In contrast to the Wilson loop case, here there are no divergences associated with the 'radial' coordinate of the AdS space. Namely, the worldsheet does not end on the boundary and hence the expression \eqref{afinal} does not suffer from divergences. We should however add the boundary term $ik\int F$ to the action in order to achieve gauge invariance. The final expression is
 \beq
\tilde S_{D5}= {N \sqrt \lambda \over 3\pi ^2} \int d\tau d\sigma \sqrt {{\rm det}\, ^\star g} \sin ^3 \theta =  T {2N \sqrt \lambda \over 3\pi} \sin ^3 \theta \int  dr \sqrt {(1 - {\omega ^2 r^2 \over {1 + {r^2 \over R^2}}})} \label{final}
\eeq
where $\theta \equiv \theta _k$ is related to $k/N$ by \eqref{result2}.

From the action \eqref{final} we read the expressions for the spin $J$ and the energy $E$ 
\beq
 J = {N \sqrt \lambda \over 3\pi ^2} \sin ^3 \theta \int  dr {{ \omega r^2 \over {1 + {r^2 \over R^2}}} \over \sqrt {(1 - {\omega ^2 r^2 \over {1 + {r^2 \over R^2}}})}}
\eeq
\beq
  E = {N \sqrt \lambda \over 3\pi ^2} \sin ^3 \theta \int  dr { 1   \over \sqrt {(1 - {\omega ^2 r^2 \over {1 + {r^2 \over R^2}}})}}
\eeq
From the above relations, as in \cite{Gubser:2002tv}, we find
\beq
E-J \equiv \Delta - J = {2N \sqrt \lambda \over 3\pi ^2} \sin ^3 \theta \log J \, 
\eeq
valid when $J \gg N \sqrt \lambda$. 

\section{Wilson loops with cusps}

Another way of deriving the result \eqref{result1} is by using the relation with cusped Wilson loops. It is well-known, due to the works by Korchemsky et.al.\cite{Korchemsky:1988si,Korchemsky:1992xv} that the cusped Wilson loop encodes the twist two operator anomalous dimension in a generic Yang-Mills theory. 

\begin{figure}[ht]
\centerline{\includegraphics[width=2.5in]{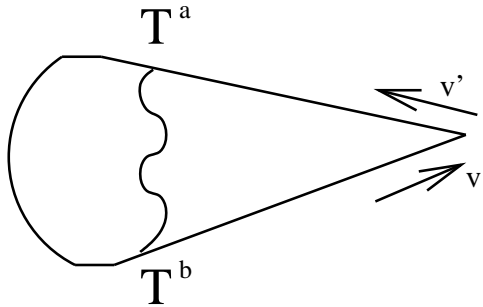}}
\caption{\footnotesize A typical perturbative contribution to a Wilson loop with a cusp. } \label{cuspfig}
\end{figure}

The relation between the cusp anomalous dimension $\Gamma _{\rm cusp}$ and the twist two operator anomalous dimension is 
\beq
\gamma  = 2 \Gamma _{\rm cusp} \log J \, , \label{cusp}
\eeq
where $\Gamma _{\rm cusp}$ is defined as
\beq
\lim _{\Psi \rightarrow \infty} \langle W_{\rm \Psi} \rangle  = \exp -(\Psi \Gamma _{\rm cusp} \log \mu / \mu_0 ) \label{wilson} \, ,
\eeq
where $W_{\rm \Psi}$ is a Wilson loop with a cusp. $\Psi$ - the cusp is defined by the relation $v v' = \cosh \Psi $, where $v$ and $v'$ are unit four-velocities of the heavy quark near the cusp, see figure \eqref{cuspfig}. $\mu$ is a UV cut-off and $\mu_0$ is an arbitrary scale. 

Due to the relations \eqref{cusp},\eqref{wilson} we can evaluate $\gamma$ by calculating cusped Wilson loops with matter in the antisymmetric representation. In fact, no new calculation is needed. Recently, by considering the DBI action of the D5 brane and comparing it to the action of the fundamental string \cite{Hartnoll:2006ib} it was found that 
\beq
{S_{\rm antisymmetric} \over S_{\rm fundamental}} = {2N \over 3\pi} \sin ^3 \theta _k \, \label{ratio}
\eeq
where $\langle W_{\rm antisymmetric}\rangle = \exp - S_{\rm antisymmetric}$,  $\langle W_{\rm fundamental} \rangle = \exp - S_{\rm fundamental}$, {\em for any shape of Wilson loop}. Thus, in particular the cusp anomalous dimension satisfies the ratio \eqref{ratio} and hence 
\beq
{\gamma _{\rm antisymmetric} \over \gamma _{\rm fundamental}} = {2N \over 3\pi} \sin ^3 \theta _k \, . \label{ratio2}
\eeq

The above re-derivation of the main result \eqref{result1},\eqref{result2} supports our claim that the spinning D5 brane computes the dimension of the boundary operator \eqref{Roperator}. The reason is as follows: in the setup of \cite{Hartnoll:2006ib} the D5-brane computes the expectation value of a Wilson loop with matter in the antisymmetric representation. Moreover, an {\em open} Wilson line generates operators of the form \eqref{Roperator} (see \cite{Belitsky:2003ys} for a recent discussion). Therefore we expect that by replacing the spinning fundamental string by a spinning D5-brane the twist two operator \eqref{Roperator} with R=fundamental should be replaced by R=antisymmetric.

\section{Discussion}

In this short note we computed the anomalous dimensions of an operator of the form \eqref{Roperator}, with $R$ the antisymmetric representation. Our result, \eqref{result1},\eqref{result2} can be approximated (within less than 3\% error for ${k\over N}=0...{1\over 2}$ ) as follows
\beq
\gamma _{\rm antisymmetric} = {N\sqrt \lambda \over \pi ^2}  \left ( \sin \pi {k\over N} - {1\over 3} (\sin \pi {k\over N})^{3\over 2} \right  ) \log J \, , \label{appresult}
\eeq

Although \eqref{appresult} is not expected to be valid in the limit where $k$ is fixed and $N
 \rightarrow \infty$ (non-interacting $k$ coincident strings), we observe that $\gamma _{\rm antisymmetric} \rightarrow k {\sqrt \lambda \over \pi}$, namely $k$ times the GKP result \cite{Gubser:2002tv}. It is clear, however, that \eqref{appresult} cannot hold at finite $k$ and finite $N$ since it is non-analytic in $1/N$ and contradicts the 't Hooft genus expansion. 

It is interesting to compare the above result \eqref{appresult} to the perturbative result \cite{Belitsky:2006en} (see \cite{Makeenko:2006ds} for a recent two-loop calculation). At order $g^2$ ('one-loop') it is clear that the dependence on $k$ is the well known 'Casimir scaling'. The reason is that at each vertex we have a factor $T^a$ and hence the contribution should be proportional to ${\rm tr}\, T^a T^b$, see figure \eqref{cuspfig}. It is known (but somewhat surprising) that the 'Casimir scaling' behaviour holds up to three loops \cite{Vogt:2004mw} \footnote{As G. Korchemsky explained in a private communication, there is no reason to believe that Casimir scaling will hold beyond three loops.}. Thus,

\beq
{\gamma _{\rm antisymmetric} \over \gamma _{\rm fundamental}} ={ k(N-k) \over N-1} \, \label{ratio-p}
\eeq

It is interesting to investigate the transition between the perturbative \eqref{ratio-p} and the non-perturbative \eqref{ratio} dependence. It is also interesting to see whether our setup can be associated with a spin chain \cite{Belitsky:2003ys,Eden:2006rx}. We postpone these issues for the future.

{\bf Acknowledgements:} I would like to thank J. Barbon, T. Hollowood, G. Korchemsky, P. Kumar, A. Naqvi and J. Ridgway for discussions . Special thanks to G. Korchemsky for a careful reading of the manuscript and to N. Drukker for comments on the first version of the paper. I am supported by the PPARC advanced fellowship award.

\end{document}